\documentclass[11pt,a4paper]{article}
\usepackage{jcappub}

\usepackage{graphicx}
\usepackage{times}%
\usepackage{natbib}
\usepackage{graphics}
\usepackage{epsfig}
\usepackage{array}
\usepackage{stfloats}
\usepackage{fixltx2e}
\usepackage{amsmath}

\newcommand{\ie}{{i.e.}}
\newcommand{\eg}{{e.g.}}
\newcommand{\gsim}{\,\lower2truept\hbox{${>\atop\hbox{\raise4truept\hbox{$\sim$}}}$}\,}

\def\eg{{\rm e.g.$\,$}}
\def\ie{{\rm i.e.$\,$}}

\newcommand{\be}{\begin{equation}}
\newcommand{\ee}{\end{equation}}
\newcommand{\bea}{\begin{eqnarray}}
\newcommand{\eea}{\end{eqnarray}}

\def\ltsima{$\; \buildrel < \over \sim \;$}
\def\simlt{\lower.5ex\hbox{\ltsima}}
\def\gtsima{$\; \buildrel > \over \sim \;$}
\def\simgt{\lower.5ex\hbox{\gtsima}}

\title{Constraints on interacting dark energy models from galaxy Rotation Curves}

\author[1,2]{Marco Baldi, }
\author[3]{Paolo Salucci}

\affiliation[1]{Excellence Cluster Universe, Boltzmannstr.~2, D-85748 Garching, Germany}
\affiliation[2]{University Observatory, Ludwig-Maximillians University Munich, Scheinerstr. 1, D-81679 Munich, Germany}
\affiliation[3]{S.I.S.S.A., Via Bonomea 265, 34136 Trieste, Italy}

\abstract{
Interacting Dark Energy models have been introduced as a possible alternative to the standard $\Lambda $CDM 
concordance cosmological scenario in order to ease the fine-tuning problems of the cosmological constant.
However, the interaction of the Dark Energy field with other massive particles in the universe induces also
an effective modification of structure formation processes, leading to a different dynamical behavior of 
density perturbations with respect to the standard scenario. In particular, high-resolution N-body simulations
have recently shown that also the structural properties of highly nonlinear objects, as e.g. their average concentration at a given mass,
could be significantly modified in the presence of an interaction between Dark Energy and Dark Matter. While a constant
interaction strength leads to less concentrated density profiles, a steep growth in time of the coupling function has
been shown to determine a large increase of halo concentrations over a wide range of masses, including the typical
halos hosting luminous spiral galaxies. This determines a substantial worsening of the ``cusp-core" tension
arising in the standard $\Lambda $CDM model and provides a direct way to constrain the form of the Dark Energy
interaction. In the present paper we make use of the outcomes of some high-resolution N-body simulations 
of a specific class of interacting Dark Energy models to compare
the predicted rotation curves of luminous spiral galaxies forming in these cosmologies against 
real observational data. Our results show how some specific interacting Dark Energy
scenarios featuring a steep growth in time of the coupling function -- which are virtually indistinguishable
from $\Lambda $CDM in the background -- cannot fit the observed rotation curves of luminous spiral galaxies and 
can therefore be ruled out only on the basis of dynamical properties of small-scale structures.
Our study is a pilot investigation of the effects of a Dark Energy interaction at small scales, and demonstrates 
how the dynamical properties of visible galaxies can in some cases provide direct constraints on the nature of Dark Energy.
}


\begin{document}
\maketitle

\section{Introduction}
\label{i}

The presently accepted standard cosmological model based on a cosmological constant $\Lambda $ and on Cold Dark Matter (CDM)
particles as the two dominant components of the Universe has proved to be extremely successful in describing the overall evolution and the structural properties
of the Cosmos on large scales. Ever improving datasets have allowed in the last decade to tightly constrain the properties of Dark Energy
(DE) at low redshifts and to show that the observed accelerated expansion of the Universe \citep[][]{Riess_etal_1998,Perlmutter_etal_1999,Schmidt_etal_1998} is driven by some field with an equation of state 
extremely close to $w_{\rm DE} = -1$ at the present epoch \citep[see e.g.][]{wmap7,Percival_etal_2001,Percival_etal_2010,Reid_etal_2010}. Furthermore, the success of the CDM paradigm 
in explaining the formation and the statistical properties of the large-scale structures observed in the Universe has been 
strongly supported by the parallel improvement of observational surveys and numerical simulations.
Nevertheless, the extreme fine-tuning of the cosmological constant value required to fit the data on the background evolution
of the Universe
poses a serious naturalness problem to the model \citep[][]{Weinberg_1988} and represents the main motivation
for the investigation of alternative scenarios. Additionally, a significant number of astrophysical observations at small scales
seem to show discrepancies with respect to the predictions of the $\Lambda $CDM model. These range from the low number
of detected luminous satellites in galactic-size CDM halos (the ``satellite problem" \cite{Navarro_Frenk_White_1995,BoylanKolchin_Bullock_Kaplinghat_2011}, but see also
e.g. \cite{Koposov_etal_2009} for a possible explanation), 
to the observed low baryon fraction in galaxy clusters \citep[][]{Ettori_2003,LaRoque_etal_2006}, the large peculiar
velocities detected in the bulk motion of galaxies \citep[][]{Watkins_etal_2009} or in systems of colliding
galaxy clusters as the ``Bullet Cluster" \citep[][]{Lee_Komatsu_2010}, to the shallow observed density profiles
of CDM halos (the so-called ``cusp-core" problem) of dwarf galaxies \citep[][]{Moore_1994,Flores_Primack_1994}, spiral galaxies \citep[][]{Salucci_Burkert_2000,Salucci_2000}, and galaxy clusters \citep[][]{Sand_etal_2002,Newman_etal_2009}.
In the present paper, we will consider the latter problem in relation to the observed dynamical properties of spiral galaxies
and discuss the impact that some specific alternative cosmological scenarios might have on this issue.
In fact, based on the shortcomings of the $\Lambda $CDM cosmology concerning both its fundamental nature and 
its observational properties at small scales, a significant number of alternative models have been proposed in recent years,
ranging from DE scenarios based on the dynamical evolution of a classical scalar field as for the case of {\em quintessence} \citep[][]{Wetterich_1988,Ratra_Peebles_1988,Ferreira_Joyce_1998} or {\em k-essence} \citep[][]{kessence}, to specific modifications of General Relativity at cosmological scales \citep[][]{Hu_Sawicki_2007}, to Warm Dark Matter (WDM) scenarios \citep[][]{Colin_AvilaReese_Valenzuela_2000,Bode_Ostriker_Turok_2001}.
A particularly interesting class of alternative models is given by interacting DE scenarios \citep[][]{Wetterich_1995,Amendola_2000}, where a direct exchange of energy-momentum between the DE field and massive particles takes place. Such models might
provide early scaling solutions for the DE density thereby alleviating the fine-tuning issues of the cosmological constant.
Additionally, they predict the existence of a fifth-force between massive particles coupled to the DE field, which determines
significant and potentially observable effects on the growth of cosmic structures \citep[see e.g.][]{Pettorino_Baccigalupi_2008,
Baldi_etal_2010,Baldi_Pettorino_2011} and is therefore likely to have a direct impact 
on the small scale failures of the $\Lambda $CDM scenario. However, the phenomenology of coupled DE models (cDE)
has been shown to be extremely diverse depending on the nature of the interaction \citep[][]{Baldi_2011a} and 
while providing a fully viable alternative to $\Lambda $CDM for a wide range of realizations,
it might
in some specific cases also significantly worsen the problems of the $\Lambda $CDM scenario on small scales while being still
in full agreement with background observables. This provides a direct way to strongly constrain the parameter space
of cDE models and represents the main focus of the present work. More specifically, we aim at quantifying the impact that a steeply
growing coupling between DE and CDM particles has on the predicted rotation curves of luminous spiral galaxies based on an NFW density profile, 
due to the change induced by the interaction on the normalization of the Concentration-Mass relation of their host DM halos. The latter effect
has been investigated through numerical N-body simulations of structure formation both for the case of a constant 
\citep[][]{Baldi_etal_2010} and a time-dependent \citep[][]{Baldi_2011a} interaction strength, and we seek here a direct
comparison of these results with observational data from galaxy rotation curves. This will allow us to rule out some 
specific realizations of the cDE scenario based on rotation curves data only.

The paper is organized as follows. In Section~\ref{models} we summarize the main features of cDE models
and we highlight the specific scenarios considered in the present work. In Section~\ref{mass_modeling}
we discuss how we model the mass distribution within spiral galaxies and how the velocity profiles are derived
based on the structural properties of the host CDM halo. In Section~\ref{constraints} we provide constraints 
on the cDE models under investigation from our sample of galaxy rotation curves, while in Section~\ref{concl}
we discuss our findings and draw our conclusions.

\section{Coupled Dark Energy models and Simulations}
\label{models}

Coupled DE models are based on the dynamical evolution of a classical scalar field $\phi $ whose energy density $\rho _{\phi }\equiv \dot{\phi }^{2}/2 + V(\phi )$
evolves in time according to the dynamic equation: 
\begin{equation}
\label{klein_gordon}
\ddot{\phi } + 3H\dot{\phi } +\frac{dV}{d\phi } = \sqrt{\frac{2}{3}}\beta _{c}(\phi ) \frac{\rho _{c}}{M_{Pl}} \,,
\end{equation}
that includes on the right-hand side a source term for the exchange of energy-momentum with a different cosmic fluid of density $\rho _{c}$.
In Eq.~\ref{klein_gordon}, an overdot indicates a derivative with respect to the cosmic time $t$, 
$H\equiv \dot{a}/a$ is the Hubble function, $M_{\rm Pl}\equiv 1/\sqrt{8\pi G}$ is the reduced Planck Mass,
$V(\phi )$ is the scalar field self-interaction potential, 
and the coupling function $\beta _{c}(\phi )$ defines the strength of the DE interaction.
For the purpose of the present paper, the fluid coupled with the DE scalar field will be always assumed to be represented by CDM particles, although
cDE models with interactions to massive neutrinos have also been proposed as a possible solution to the cosmic coincidence problem \citep[see \eg][]{Amendola_Baldi_Wetterich_2008,Baldi_etal_2011a}.
In order to preserve General Covariance, Eq.~\ref{klein_gordon} requires the presence of an analogous interaction term also in the CDM continuity equation,
which reads:
\begin{equation}
\label{continuity_cdm}
\dot{\rho }_{c} + 3H\rho _{c} = -\sqrt{\frac{2}{3}}\beta _{c}(\phi )\frac{\rho _{c}\dot{\phi }}{M_{Pl}} \,,
\end{equation}
while the remaining continuity equations for the baryonic and relativistic components of the universe are unaffected as the coupling involves only DE and CDM:
\begin{eqnarray}
\dot{\rho }_{b} + 3H\rho _{b} &=& 0 \,, \\
\dot{\rho }_{r} + 4H\rho _{r} &=& 0\,.
\end{eqnarray}
As a consequence of the interaction, and assuming the conservation of the CDM particle number, Eq.~\ref{continuity_cdm} implies a time evolution of the 
CDM particle mass due to the dynamic nature of the DE scalar field, according to the equation:
\begin{equation}
\label{mass_variation}
m_{c}(a) = m_{c}(a_{0})e^{-\sqrt{2/3}\int \beta _{c}(\phi )d\phi /M_{Pl}} \,,
\end{equation}
where $a_{0}$ is the cosmic scale factor at the present time.
The cosmological dynamics of the system allows for scaling solutions during matter domination that feature an Early Dark Energy component, thereby alleviating
the fine-tuning problems of the cosmological constant. Such scaling solutions have been identified analytically for the case of constant coupling functions $\beta _{c}(\phi )=\beta _{c} = {\rm const.}$, and numerically for the case of variable couplings $\beta _{c}(\phi )$ considered in the present work , and 
are known as the ``$\phi $-{\em Matter Dominated Epoch}" \citep[or $\phi $MDE, see ][]{Amendola_2000} and the ``Growing-$\phi $-{\em Matter Dominated Epoch}"
\citep[or G-$\phi $MDE, see ][]{Baldi_2011a}, respectively.
However, if the coupling exhibits a very steep growth in time (as for the case of the specific models discussed in this work) the
early DE scaling is absent and the model is therefore indistinguishable from $\Lambda $CDM at the background level \citep[see again ][]{Baldi_2011a}.

At the linear perturbations level, the growth of density perturbations is modified by the interaction between DE and CDM through the appearance of two new terms that directly depend on the DE-CDM coupling in the 
perturbed linear equations. The dynamic equation for linear CDM density perturbations
in the Newtonian gauge and in Fourier space is in fact modified from its standard form and reads \citep[see \eg][for a full derivation of the linear perturbation equations in cDE models]{Amendola_2000,Amendola_2004,Pettorino_Baccigalupi_2008,Baldi_etal_2010}:
\begin{equation}
\label{gf_cdm}
\ddot{\delta }_{c} = -2H\left[ 1 - \beta _{c}(\phi )\frac{\dot{\phi }}{H\sqrt{6}}\right] \dot{\delta }_{c} + 4\pi G \left[ \rho _{b}\delta _{b} + \rho _{c}\delta _{c}\left( 1 + \frac{4}{3}\beta_{c} ^{2}(\phi )\right) \right] \,, 
\end{equation}
where the overdensity $\delta _{b,c}$ of baryons and DM is defined as $\delta _{b,c}\equiv \delta \rho _{b,c}/\rho_{b,c}$\,.
The additional contribution appearing in the first term on the right hand side of Eq.~(\ref{gf_cdm}) is an extra friction associated with
momentum conservation in cDE models \citep[see \eg ][for a discussion on the effects of the friction term]{Baldi_2011b} while
the additional term in the second squared bracket
includes the effect of the fifth-force mediated by the DE scalar field for CDM density perturbations. Qualitatively, while the fifth-force term is always attractive
and accelerates the growth of density perturbations for any value of the coupling $\beta _{c}$ and for any dynamic evolution of the DE scalar field $\phi $,
the friction term enhances the growth only when $\beta _{c}\dot{\phi } > 0$, while it slows down the evolution of perturbations when $\beta _{c}\dot{\phi } <0$.
Although in the present work we will be interested in specific cDE models for which the former condition always holds, a change in sing of the friction term
can have particularly interesting effects on structure formation processes, especially at high redshifts \citep[][]{Baldi_2011c,Tarrant_etal_2011}.
Furthermore, even for the case of cDE models with always positive values of $\beta _{c}(\phi )\dot{\phi }$, the friction term has been shown to determine a
suppression of the growth of density perturbations once these become nonlinear, due to the presence of non-radial velocities \citep[][]{Baldi_etal_2010,Baldi_2011b}.

In the present work, we will focus on cDE models with an exponential self-interaction potential and with variable couplings in the form of a power of the
cosmic scale factor $a$:
\begin{equation}
\beta _{c}(\phi (a)) \equiv \beta _{0}a^{\beta _{1}} 
\end{equation}
that have been proposed and studied by \citep[][]{Baldi_2011a}. In particular, we will consider two of the models discussed in \citep[][]{Baldi_2011a}, featuring
a rapidly growing coupling function with $(\beta _{0},\beta _{1}) = (0.5,2)$ and $(0.75,3)$, named ``EXP010a2" and ``EXP015a3", respectively. Such models have been shown to be fully consistent
with present cosmological data on the background expansion history and cluster number counts,  while determining a significant increase of the 
small-scale matter power only at very low redshifts. In the present paper, we are seeking for observational constraints on this specific class of models
based only on the observation of rotation curves of luminous spiral galaxies, without invoking any additional observational data.

\begin{figure}
\includegraphics[scale=0.5]{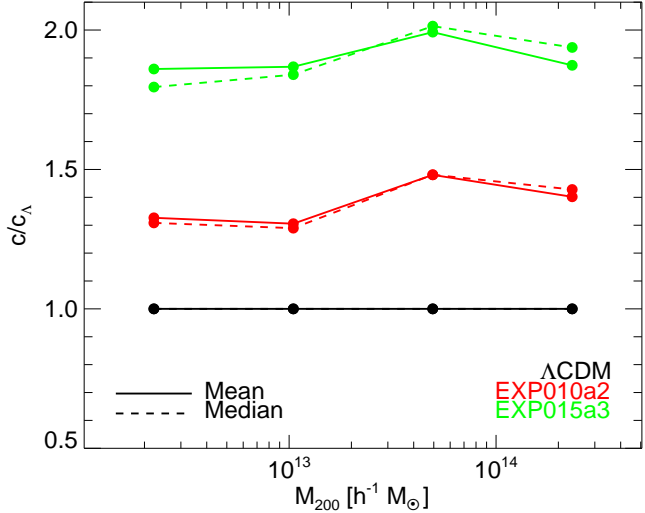}
\caption{The ratio of the halo concentration in cDE models with respect to the standard $\Lambda $CDM scenario as a function of the halo virial mass $M_{200}$, as obtained from the N-body simulations of \cite{Baldi_2011a}. The curves represent the Mean (solid) and the Median (dashed) concentration within each of the four logarithmic mass bins in which the sample has been subdivided.}
\label{fig:concentration}
\end{figure}

We will base our analysis on the numerical results of \citep[][]{Baldi_2011a}, that made use of the specific modification of \cite{Baldi_etal_2010} of the
widely used parallel N-body code {\small GADGET} \citep[][]{gadget-2} to run high-resolution hydrodynamical N-body simulations of a series of cDE models with variable coupling, 
including the EXP010a2 and EXP015a3 models under investigation here. Such modified numerical code includes all the specific effects related to the DE-CDM interaction that have been 
briefly discussed above, and allows to follow the evolution of a cosmological volume of the universe in the context of different models with a
common normalization of the basic cosmological parameters, and to compare the individual and statistical properties of collapsed objects forming in each
scenario. In particular, we will consider here the results of hydrodynamical N-body simulations within a periodic cosmological box of $80$ comoving Mpc$/h$
aside filled with $512^{3}$ CDM and $512^{3}$ gas particles, for a mass resolution of $2.41 \times 10^{8}$ M$_{\odot }/h$ and $4.82 \times 10^{7}$ M$_{\odot }/h$,
respectively, and a spatial resolution of $3.5$ kpc$/h$. By comparing the outcomes of such simulations, \cite{Baldi_2011a} found that -- differently
from standard cDE scenarios with constant coupling -- variable coupling models might determine in some specific realizations a significant increase of the normalization of the ``concentration {\em vs.}
Mass" (c-M) relation for collapsed virialized objects (see Fig.~\ref{fig:concentration}). 
In particular, this happens for models where the coupling steeply grows at low redshifts as it is the case for
couplings scaling as a positive power of the scale factor, due to the absence in such models of a G-$\phi $-MDE phase \citep[see ][for a detailed discussion
of this effect]{Baldi_2011a}. An overall increase of the normalization in the c-M relation might have a significant impact on the predicted rotation curves for luminous spiral galaxies
and provide a powerful way to strongly constrain the parameter space for this class of cDE models.  More specifically, an increase by  a factor of $\sim 1.85$ 
\citep[as predicted by][for the model EXP015a3]{Baldi_2011a} of the expected average concentration of CDM halos in the mass range 
of a few $\times 10^{12}$ M$_{\odot }/h$ will clearly exacerbate the marginal tension of observed rotation curves with the predicted radial mass distribution
of a Navarro-Frenk-White \citep[NFW][]{Navarro_Frenk_White_1995} density profile in $\Lambda $CDM, known as the ``cusp-core" problem, thereby providing a direct way to
constrain the model using only galaxy rotation curves data and without resorting on any additional observational dataset.

In this paper we will adopt the outcomes of \citep[][]{Baldi_2011a}, where simulations were normalized in order to have the same amplitude of linear density perturbations
at $z=0$, \ie all cosmological models were normalized to the same value of $\sigma _{8}(z=0)$. Recently, a new set of very large N-body simulations for cDE
models \citep[the {\small CoDECS} project, see][]{CoDECS} has been released, featuring a different choice for the power spectrum normalization with a common
amplitude of density perturbations at the redshift of CMB, $z_{\rm CMB}\approx 1100$. The choice to consider the numerical results of \citep[][]{Baldi_2011a} instead of
\citep[][]{CoDECS} for our analysis should be considered ``conservative" as a high-$z$ normalization of density perturbations is expected to result in an even stronger increase of halo
concentration at a fixed mass for the same cosmological model.

\section{Mass modelling the Spiral's Rotation Curves}
\label{mass_modeling}

In spiral galaxies the presence of a large amount of unseen matter, distributed very differently from the stellar and HI gas disks, is well established from their non-keplerian rotation curves (\cite{Rubin_Ford_Thonnard_1980,Bosma_1981}, an innovative review on this issue can be found in \cite{Salucci_FrigerioMartins_Lapi_2011}). A massive dark component, becoming progressively more important at increasing radii and decreasing galaxy luminosity is present in any galaxy \citep[][]{Persic_Salucci_1988}. This evidence relies on the fact that rotation curves (RCs) of spiral galaxies suitably measure their total gravitational potential \citep[see e.g.][]{Ratnam_Salucci_2000,Yegorova_Salucci_2007}. More precisely, high-quality RCs measure the spiral's circular velocity $V(r) $, i.e. the rotational equilibrium velocity implied by their underlying mass distribution. In each galaxy, the former is related to the total gravitational potential $\phi _{\rm tot}$,
which is the sum of four distinct mass components: a spherical stellar bulge ($B$), a Dark Matter halo ($DM$), a stellar disk ($*$) and a gaseous disc ($HI$). 
Such relation follows the equation:
\begin{equation}
\label{RCs}
V^2(r)=V^2_B + V^2_{DM}+V^2_*+V^2_{HI}= r\frac{d}{dr}\phi_{tot} \,.
\end{equation}
where $\phi_{\rm tot}=\phi_{B}+ \phi_{DM}+\phi_*+\phi_{HI}$.
It is therefore possible to equal the observed circular velocity, i.e the LHS of Eq.~\ref{RCs}, with the velocity model given by the RHS of Eq.~\ref{RCs},
thereby inferring the radial profile of the total gravitational potential. More specifically, this is obtained by means of the Poisson equation from the surface/spatial densities of the various mass components.

\subsection{The Baryonic contribution}

The surface density of the stellar disk $\Sigma_*(r)$ is obtained from the observed surface brightness, once we assume a radially constant stellar mass-to-light ratio \citep[see][]{Portinari_Salucci_2010}. It is generally found \citep[see e.g.][]{Freeman_1970} that this quantity follows the relation:
\begin{equation}
\Sigma_{*}(r)=\frac{M_{D}}{2 \pi R_{D}^{2}}\: e^{-r/R_{D}}
\end{equation}
where $M_D$ is the disk mass and $R_D$ is the disk scale length; therefore, for the RC we get a contribution:
\begin{equation}
V_{*}^{2}(r)=\frac{G M_{D}}{2R_{D}} x^{2}B\left(\frac{x}{2}\right)
\end{equation}
from the stellar disk, where $x\equiv r/R_{D}$, $G$ is the gravitational constant, and $B=I_{0}K_{0}-I_{1}K_{1}$ is a combination of Bessel functions.

The contribution of the gaseous disk to the RC, $V^2_{HI}$, can be directly derived from the HI surface brightness; however, for the high-luminosity spirals we consider here, this contribution is always negligible \citep[see e.g.][]{Bosma_1981}. Moreover, although spirals can have a non-negligible bulge velocity component, 
this results virtually indistinguishable from that of the exponential disk \citep[see][]{Persic_Salucci_Ashman_1993} for the determination of the DM structural mass parameters via RC fitting.

\subsection {The phenomenology of RCs of Spirals}

The systematical study of the spiral RCs has highlighted their universal character, with the galaxy magnitude $M_{I}$ playing the role of an additional characteristic parameter
identifying specific galaxies along a universal behavior of the RC \citep[][]{Persic_Salucci_Stel_1996,Salucci_etal_2007}. This has led to the concept of ``Universal Rotation Curve'' (URC), i.e. a function $V_{URC}(r/R_D,M_{I})$ of radius (in units of the disk length scale, $r/R_D$) and of galaxy magnitude ($M_{I}$) that well reproduces the RC of any object of known $M_{I}$ and $R_D$ \citep[see][]{Persic_Salucci_Stel_1996}. In detail, \cite{Persic_Salucci_Stel_1996} assigned each of their 900 individual RCs of spirals extended out to $R_{opt} \simeq 3 R_D $ to the corresponding magnitude bin among the 11 bins in which the whole spiral $I$-band magnitude range $M_{I}\in \left[ -24.3,-16.3\right] $ was divided.
The RCs of the objects in each luminosity bin were finally averaged in radial bins of width $0.1 R_{opt} $. This led to a family of synthetic coadded RCs 
$V_{coadd}(r/R_D,M_I)$ \citep[see e.g. Fig.~1 of][]{Persic_Salucci_Stel_1996}. A similar result was obtained by \cite{Catinella_Giovanelli_Haynes_2006}. These coadded RCs result regular, smooth and with a very small intrinsic variance. Additional data, including very extended RCs and the virial velocities $V_{vir}\equiv (G M_{vir}/R_{vir})^{1/2}$ \citep[see][]{Shankar_etal_2006}, allowed to extend the coadded RC out to the galaxies virial radii \citep[][]{Shankar_etal_2006,Salucci_etal_2007}.
With such a general prescription, all kinematical data were fit by the same analytical function $V_{URC}(r/R_D,M_{I})$ \citep[see again][]{Salucci_etal_2007}. 
Further evidence that the URC does represent the general RC of any spiral of any luminosity was provided by \citep[][]{Catinella_Giovanelli_Haynes_2006}.

\begin{figure}
\includegraphics[scale=0.7]{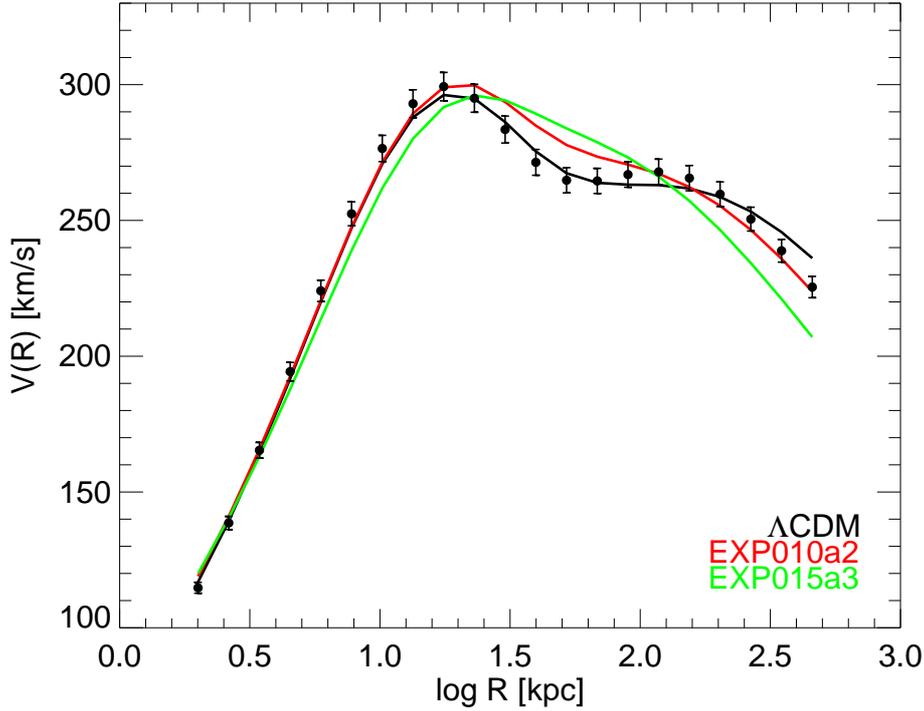}
\caption{Circular rotation velocity for the case of a standard $\Lambda $CDM concentration {\em vs.} Mass relation (black solid line) and for the two cDE models under investigation, EXP010a2 (red solid line) and EXP015a3 (green solid line), compared to the URC at $M_{I} = -23$ (black points with error bars).}
\label{RC}
\end{figure}

In this paper we are concerned with RCs of high-luminosity spirals, as their DM halos correspond to the smallest objects 
for which the simulations by \cite{Baldi_2011a} provide a reliable estimation of the halo structural parameters. As a matter of fact, we aim to compare the URC relative to the top-luminosity spiral  objects -- i.e. to objects with $M_I \leq -23$ \citep[see the two last panels in Fig.~1 of][]{Persic_Salucci_Stel_1996} -- with the prediction of some specific mass models.
The former is plotted in Fig.~\ref{RC} as black filled circles, with error bars that include also the variance detected object by object in the 56 RCs of \cite{Persic_Salucci_Stel_1996} and the 30 of \cite{Yegorova_etal_2011}, i.e. of the available samples of high luminosity spiral RCs. More specifically, the data points in Fig.~\ref{RC} show a {\it typical} circular velocity profile of a ``very big" spiral, $M_I =-23.0$ and consequently with $R_D=7.3 $ kpc, $V(R_{opt})=235 km/s $ \citep[see][although no result in the present paper depends on these numbers and  exactly the same result would arise if we had used anyone of the available  individual RC of  high luminosity spira as all of them  are, for  the present aim, indistinguishable from the adopted one]{Salucci_etal_2007}.
Obviously, any successful mass model must be capable to reproduce these data.

\subsection{The velocity profile in $\Lambda$CDM halos}

It is well known that numerical simulations performed in the $\Lambda$CDM scenario lead to virialized halos of Dark Matter with a very specific density and velocity profile, called the NFW profile \citep[][]{Navarro_Frenk_White_1995}:
\begin{equation}
\rho_{NFW}(r) = \frac{\rho_s}{(r/r_s)\left(1+r/r_s\right)^2}\,.
\label{eq:nfw}
\end{equation}
In Eq.~(\ref{eq:nfw}) $r_s$ is a characteristic scale radius and $\rho_s$ is the corresponding density. We remind that for the virial radius $R_{vir}$ and halo mass $M_{vir}$ and the mean universal density $\rho_u$ we have: 
$M_{vir} \simeq 100 \rho_u R_{vir}^3$. 
Furthermore, $r_s$ and $R_{vir}$ are found to be related within a reasonable scatter according to the equation \citep[][]{Klypin_TrujilloGomez_Primack_2011}:
\begin{equation}
\label{concentration_eq}
c = R_{vir} / r_s \simeq 9.7 \left( \frac{M_{vir}}{10^{12}M_{\odot}} \right)^{-0.13}\,.
\end{equation}
Although other mildly different density profiles and/or $c$ vs. $M_{vir} $ relations have been proposed, adopting Eq.~(\ref{concentration_eq}) as the standard $\Lambda $CDM concentration does not affect the outcomes of the present work.

Since they emerged in N-body simulations, the cuspy NFW density profiles have been in tension with those detected around dwarf spirals \citep[][]{Moore_1994}, 
giving rise to the so called ``cusp-core" problem. Later, e.g. in \cite{Salucci_Walter_Borriello_2003} and \citep{Gentile_etal_2004}, a suitable investigation of a number of  proper test-cases was performed by means of a careful analysis of 2D, high-quality, extended, regular, and symmetric RCs for spiral galaxies. 
It is well known that in these objects the NFW halo predictions and kinematical data are found in strong disagreement in several aspects. In fact, the stellar disk + NFW halo mass model (NFWD)
{\it i)} fits very poorly a significant number of RCs; {\it ii) } often requires an implausibly low stellar mass-to-light ratio and/or {\it iii)} an unphysical high halo mass and/or {\it iv)} an implausible value for the concentration parameter $c$ \citep[see e.g.][]{Spekkens_Giovanelli_Haynes_2005,Gentile_etal_2004,Gentile_etal_2005,Gentile_etal_2007,Spano_etal_2007,deBlok_etal_2008,deNaray_McGaugh_deBlok_2008}.
Up to now, the failure of the NFW velocity profiles is very evident in low luminosity galaxies, as their steep RCs cannot be accounted for by a NFWD velocity profile. Noticeably, instead, for the most luminous objects as large spiral galaxies, their flattish RCs can be still (marginally) reproduced also by a NFWD mass model. As an example, the RCs of the luminous spirals MW and M31 can be well fitted also by a NFWD mass model \citep[][]{Chemin_Carignan_Foster_2009}.
\ \\

The $\Lambda $CDM scenario might generically account for the ``cusp-core" tension described above if some physical process, related to the galaxy formation process itself, erased the previously formed cosmological cusp. A rather common approach is therefore to assume a $\Lambda $CDM scenario and to explain the observed RCs postulating that some process has transformed the originally sharp halo density profiles into the observed shallow ones. Consequently, any process or scenario in which $\Lambda $CDM halos get a {\it more} pronounced cusp with respect to the standard case should be disregarded.
In fact, given the present possible tension between $\Lambda $CDM halos predictions and actual observations, it is possible to rule out or tightly constrain any scenario that further worsens the discrepancy. In other words, any reasonable process beyond the adiabatic formation of CDM halos in $\Lambda $CDM we want to invoke must erase the central cusp and not make it more pronounced.

\section {Constraining cDE models with RCs of High-Luminosity Spirals}
\label{constraints}
\begin{figure}
\label{chi2}
\includegraphics[scale=0.5]{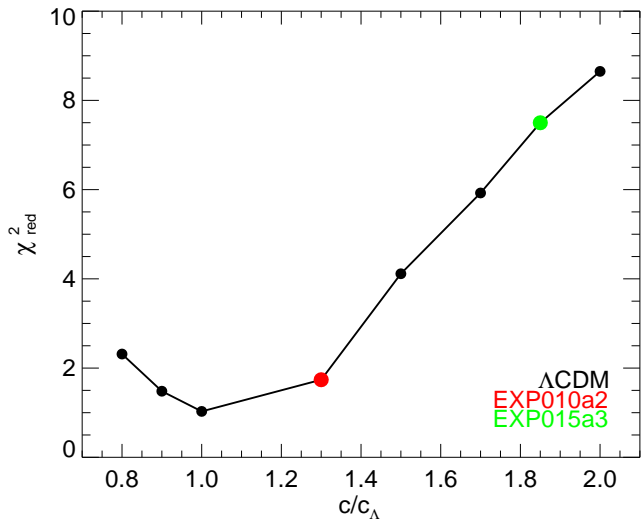}
\caption{Values of $\chi _{\rm red}^{2}$of the best-fit mass models for different assumptions on the concentration parameter $c/c_{\Lambda }$. Standard NFW models correspond to $ c/c_{\Lambda}=1$. The $\chi _{\rm red}^{2}$ values for the two cDE models
discussed in the present work are shown by a red (EXP010a2) and a green (EXP015a3) dot. The latter model provides a very poor fit
to the data, as shown by the large $\chi _{\rm red}^{2}$ value.}
\end{figure}

In Fig.~\ref{RC} we plot (as black points) the synthetic RC of spiral galaxies of $M_I \sim -23$ from 1 kpc, their typical innermost radius with reliable individual measurements, to 330 kpc (their typical virial radius). It is worth to recall here that the cored halo + Freeman Disk model (not discussed here) perfectly fits these data by passing all the data points within their error bars \citep[][]{Salucci_etal_2007}.

It is also not surprising that the NFWD model with the standard concentration relation (\ref{concentration_eq}) fits the same data in a quite satisfactory way (black line in Fig.~\ref{RC} and black filled circle in Fig.~\ref{chi2} where the resulting reduced chi-squared $\chi^{2}_{\rm red}$ is shown). In fact, so far in the literature there are no cases in which the flattish individual RCs of high-luminosity spiral galaxies can be unambiguously modeled exclusively by a cored DM distribution as it occurs for the steeper RCs of lower luminosity objects.

Let us now investigate the case -- as for the cDE scenarios with a steep growth in time of the coupling function $\beta _{c}$ \citep[][]{Baldi_2011a} -- 
in which the emerging dark matter halos have a NFW density profile with a value for the concentration parameter significantly higher with respect to 
what prescribed by Eq.~(\ref{concentration_eq}) for halos of this mass
For instance, we can consider concentration values as found in N-body simulations for the EXP010a2 and EXP015a3 models
investigated in the present work (see Fig.~\ref{concentrations}), which at masses of the order of a few $\times 10^{12}M_{\odot }$
are given by:
\begin{eqnarray}
{\rm EXP010a2}\qquad c &\approx &1.3 \cdot c_{\Lambda }\,, \\
{\rm EXP015a3}\qquad c &\approx & 1.85 \cdot c_{\Lambda }\,,
\end{eqnarray}
where $c_{\Lambda }$ is the standard value of the concentration predicted by $\Lambda $CDM simulations and consistent with Eq.~\ref{concentration_eq}.

With these values of the concentration parameter, we can now attempt fitting the reference RCs of high-luminosity spirals shown in  Fig.~\ref{RC} with a NFWD mass model for NFW DM halos of different concentration, and test the quality of the fit
with the reduced chi-squared $\chi _{\rm red}^{2}$ defined as:
\begin{equation}
\chi _{\rm red}^{2} \equiv \frac{1}{N-2} \sum_{1}^{N} \frac{V_{\rm URC}-V_{\rm mod}}{0.0175 \cdot V_{\rm URC}^{2}} \,,
\end{equation}
where N=19, the quantity in the denominator is the observational error of $V_{\rm URC}$ and $V_{\rm mod}$ is the velocity model under investigation.
For a standard $\Lambda $CDM model, corresponding to $c/c_{\Lambda }=1$, the best fit parameters are 
$M_{vir }=1.1 \times 10 ^{13} M_\odot$, $M_D=5.1 \times 10^{11} M_\odot$ with a reduced chi-squared of $\chi^2_{\rm red}\sim 1$.

The fit appears instead clearly unsuccessful (see the green line in Fig.~\ref{RC} and the green point in Fig.~\ref{chi2}) 
for the most extreme cDE case EXP015a3, corresponding to the parameters $(\beta _{0},\beta _{1})=(0.75,3)$. 
The best fit values of the parameters for this model are $M_{vir }=6 \times 10 ^{12} M_\odot$, $M_D=2.6 \times 10^{11} M_\odot$, but the fit results extremely poor with a reduced chi-squared of $\chi^2_{\rm red}\sim 7.5$. 
On the other hand, the fit is not significantly worse than in the standard $\Lambda $CDM case for the other, less extreme cDE scenario
discussed in this work, EXP010a2, with $(\beta _{0},\beta _{1})=(0.5,2)$, for which the best fit parameters are $M_{vir}=8.7 \times 10 ^{12} M_\odot$, $M_D=4.1 \times 10^{11} M_\odot$ with a $\chi^2_{\rm red}\sim 1.7$.
Let us stress here that the fitting uncertainty in the the two structural parameters is large but not relevant here. In the former case, 
no model can fit the data in a satisfactory way.

More in general, let us consider DM halos with a NFW profile but with a c-M relation different from the standard $\Lambda $CDM case,
i.e. $c/c_{\Lambda }$ ranging from $c/c_{\Lambda } = 0.8$, leading to a marginally shallower density profile, to $c/c_{\Lambda } = 2$, i.e. a significantly more concentrated DM halo.
We attempt in these cases a NFWD modeling of the reference RC. The result is clear (see Fig.~\ref{chi2}): at a value of $c/c_{\Lambda } \gsim 1.8$ the fit becomes very problematic and even flat RCs rule out a NFW halo with such a high concentration.

\section{Discussion and Conclusions}
\label{concl}

In the $\Lambda $CDM framework, DM virialized halos are characterized by a cuspy density distribution that is certainly not seen in the halos around dwarf spiral galaxies and that might be absent also in high-luminosity spiral galaxies. 
Some alternative cosmological models, motivated by the need to address the fundamental problems of the cosmological
constant, as e.g. some specific models of interaction between Dark Energy and Dark Matter
can modify the typical concentration of DM halos as a function f halo mass, both reducing it or increasing it
depending on the specific form of the DE self-interaction potential and of the coupling function.
However, while the reduction of halo concentration, when it occurs, is in general quite modest for presently
viable cDE models, some specific scenarios can
lead to DM halos that are significantly more concentrated than the standard $\Lambda $CDM ones. As a consequence of this, also the rotation curves of high-luminosity spiral galaxies as obtained from the Universal Rotation Curve or from the kinematics of numerous suitable objects, cannot be reproduced in a satisfactory way by a mass model that accounts for the effect of enhanced concentration in interacting DE cosmologies.

In the present work, we have investigated the impact that these specific realizations of the interacting DE scenario have on the rotation curves of luminous spiral galaxies.
More specifically, we have restricted our analysis to a particular class of coupled DE scenarios characterized by a steep growth
in time of the interaction strength, parametrized as a positive power of the cosmic scale factor. Such models have been shown
to be in full agreement with present constraints on the background evolution of the Universe, as they do not feature the typical
early DE scaling of other coupled DE scenarios like e.g. models with a constant or a more slowly evolving coupling function. The 
background evolution of the models investigated in this work is in fact practically indistinguishable from the standard $\Lambda $CDM
cosmology.
Furthermore, the impact of these alternative cosmologies on the statistical properties of large scale structures is confined at very low redshifts, which makes it problematic to test the models via e.g. weak lensing measurements. 

However, as mentioned above, a series of high-resolution hydrodynamical N-body simulations for this specific class of interacting DE cosmologies have 
recently shown that the fast growth of the coupling strength determines a significant increase of the normalization of the concentration {\em vs.} Mass relation for CDM halos at low redshifts, thereby allowing for a direct test of the models through dynamical probes
in collapsed structures. In the present study we have therefore considered the effect that an average higher halo concentration 
at halo masses in the range of luminous spirals  -- consistent with the results of N-body simulations for two specific realizations of coupled DE models -- determines on the rotation curve of the galaxy, and compared the inferred circular velocity with real data,
namely with the Universal Rotation Curve that for our purpose well represents the typical circular velocity of spirals. 

While the standard $\Lambda $CDM value of the halo concentration provides a marginally acceptable fit to the data, the increase
of the concentration due to the interaction between DE and DM for the specific models investigated in this work determines
a significantly worse fit and exacerbates the problem of the central Dark Matter cusps in galactic halos. 
In particular, our results show that for the most extreme of our coupled DE scenarios the
best fit mass modeling has a reduced chi-squared of $\chi _{\rm red}^{2} \approx 7.5$, 
thereby allowing to rule out the model.
The other model under investigation, which features a smaller present value and a shallower growth in time of the coupling function,
shows instead a still marginally acceptable fit to the data.

We can therefore conclude that the nonlinear effects of an interaction between DE and CDM provide a direct
way to constrain the evolution of the coupling function, and that the parameter space of coupled DE models can be
constrained by means of direct observations of the dynamical properties of galaxies. Furthermore, since the problem of
the central cuspiness of DM halos is mostly severe for faint and low-mass objects, we argue that a detailed investigation
of the effects of coupled DE models on the halo concentration at a lower mass range than allowed by the present
resolution of our simulations might result in tighter constraints on the coupled DE parameter space. 
It is also important to stress here that a large fraction of the parameter space of coupled DE cosmologies does not result
at all in an enhancement of halo concentrations, with a large number of models even showing a slight decrease of the 
concentrations over a wide range of masses. These realizations of the coupled DE scenario, not discussed in the present work,
are clearly not affected by our constraints and should still be considered as viable alternatives to the $\Lambda $CDM paradigm.
Nevertheless, our results show for the first time that it is possible to constrain the properties of Dark Energy also through
the dynamical features of luminous galaxies.

\section*{Acknowledgments}
MB acknowledges support by 
the DFG Cluster of Excellence ``Origin and Structure of the Universe''
and by the TRR33 Transregio Collaborative Research
Network on the ``Dark Universe''. MB also wants to acknowledge SISSA
and the HPC-Europa2 visiting grant nr. 589 for financial support 
during his visits at SISSA.

\bibliographystyle{JHEP}
\bibliography{baldi_bibliography}

\label{lastpage}

\end{document}